\documentclass[
 reprint,
 nofootinbib,
 amsmath,amssymb,
 aps,
 prl
]{revtex4-2}
\usepackage{hyperref}
\usepackage{graphicx}
\usepackage{dcolumn}
\usepackage{bm}
\usepackage{tikz}
\usepackage{quantikz}
\usepackage{adjustbox}
\usepackage{braket}
\usepackage{amsmath}
\usepackage{amssymb}
\usetikzlibrary{quantikz2}

\usepackage{xspace}
\newcommand{\idest}{i.e.,\xspace}

\newcommand{\pars}[1]{\left( #1 \right)}

\newcommand{\myket}[1]{\left| #1 \right \rangle}

\begin{document}

\preprint{APS/123-QED}

\title{Comment on ``Efficient quantum state preparation with Walsh series"}

\author{Riccardo Pellini}
\email{riccardo.pellini@polimi.it}
\author{Maurizio {Ferrari Dacrema}}
 \email{maurizio.ferrari@polimi.it}
\affiliation{
 Politecnico di Milano, Milan, Italy
}

\date{\today}

\begin{abstract}
In this paper, we discuss the Walsh Series Loader (WSL) algorithm, proposed by \citet{Zylberman2024}. In particular, we observe that the paper does not describe how to implement the term of order zero of the operator WSL is based on.
While this does not affect the theoretical validity of WSL, it poses obstacles for practitioners aiming to use WSL because, as we show in our experiments, an incorrect implementation leads to states with very poor fidelity.
In this paper, we describe how to implement the full quantum circuit required by WSL, including the term of order zero, releasing our source code online, and show that the algorithm works correctly. Finally, we  empirically show how, as was theoretically demonstrated in the original article, truncated WSL does not prepare well highly entangled states. 
\end{abstract}

\maketitle

\section{\label{sec:introduction}Introduction}
State preparation is a key task to tackle in order to map classical data onto quantum systems. Such task consists in loading a complex-valued vector $x$ of dimension $N=2^n$, with $n \in \mathbb{N}$, into the amplitudes of a quantum register $\myket{x}$ of $n$ qubits. This encoding is called \emph{amplitude encoding}. In particular, if $x = \left(
   x_0  ,  x_1 , \cdots , x_{N-1}
\right)$, the state to prepare is given by:
\begin{equation}
    \myket{x} = \frac{1}{\sqrt{\sum_{i=1}^{N-1}|x_i|^2}}\sum_{i=0}^{N-1} x_i \myket{i}
\end{equation}
where the state $\myket{i}$ identifies the eigenstate given by the binary representation of the integer $i$.

The Walsh Series Loader (WSL), developed by \citet{Zylberman2024}, is an approximate method of state preparation. WSL is based on the expansion of a real-valued function $f: [0,1) \to \mathbb{R}$ with a generalized Fourier series, called \emph{Walsh series}. 

The contributions of this paper are the following:
\begin{itemize}
    \item \textbf{The implementation of the WSL algorithm is incomplete:}
    While the original paper describes how to implement the operators for the Walsh terms of order $\geq 1$, it does not mention that the term of order zero is a special case that requires a different implementation;
    \item \textbf{The missing description limits reproducibility:} An incomplete implementation of WSL in which the term of order zero is missing leads to very poor results compared to those reported in the original paper, highlighting its importance;   
    \item \textbf{We provide a correct WSL circuit:} we identify how the missing Walsh term can be added into the quantum circuit that implements WSL and show empirically that the method works correctly.\footnote{We release online an implementation of the algorithm at \url{https://github.com/qcpolimi/WSL}}
\end{itemize}

\section{Walsh Series Loader}
\label{sec:wsl}
In this section, we provide a brief description of a correct implementation of the Walsh Series Loader (WSL).

\subsection{Walsh Series, Walsh Functions and Walsh Transform}
\label{sec:walsh_objects}
Given the discrete real function $f: D \to \mathbb{R}$ defined over $N=2^n$ points,\footnote{While $f$ is defined over the interval $[0,1)$, since it has to be mapped into a quantum state we assume it is discrete. The cardinality of the domain $D$ must be a power of 2 and we use directly integer numbers identify the values of $f$.} with $D = \left\{ 0, 1, 2, \cdots, N-1 \right\}$, we can expand such discrete function as a \emph{Walsh Series}, \idest a finite sum of $N$ \emph{Walsh terms}.
Each term is the product between a coefficient $a_h$, called \emph{Walsh transform}, and a function $w_h(k)$ called \emph{Walsh function}, where $h$ represents their order.
The Walsh Series is defined as:\footnote{Since $f$ is a discrete function Equation \ref{eq:walsh_series} identifies a full Walsh series; if $f$ was continuous it would identify a truncated one.}
\begin{equation}\label{eq:walsh_series}
    f(k) = \sum_{h=0}^{N-1}a_h w_h(k)
\end{equation}
Function $f$ can be approximated by using only $M<N$ Walsh terms, where $M$ can be determined based on the accepted error $\epsilon_1 \in (0,1)$. In particular, $M = 2^m$ with $m=\left\lceil \log_2\left( \frac{1}{\epsilon_1} \right) \right\rceil$. Since $f$ is discrete, M must be a power of two to ensure the points are contained in $D$.

The Walsh function $w_h(k)$ is defined as follows:\footnote{In the original paper \cite{Zylberman2024}, the definition is slightly different because $k_{j}$ was defined based on a dyadic expansion. Here we simplify the notation using for both $h$ and $k$ a binary expansion.
}
\begin{equation}
    w_h(k) = (-1)^{\sum_{j=0}^{n-1}h_{j}k_{n-j}}
\end{equation}
where $h_{j}$ and $k_{j}$ represent the $j$-th least significant bit in the binary representation of $h$ and $k$.

The Walsh transform $a_h$ over $M$ terms is defined as:
\begin{equation}\label{eq:Walsh_transform}
    a_h = \frac{1}{M}\sum_{k=0}^{M-1}w_h(k)f(k)
\end{equation}

\subsection{Walsh Operators}
\label{sec:walsh_operator}
The WSL algorithm is based on an operator $\hat{U}_{f}$ proposed by \citet{Welch2014} to represent Walsh terms, which is applied as a controlled unitary.
In this section we discuss the non-controlled unitary:
\begin{equation}
\label{eq:walsh_unitary}
\hat{U}_{f} 
    = \prod^{M-1}_{h=0} e^{-ia_h\hat{w}_h}
    = \prod^{M-1}_{h=0} \hat{W}_h(a_h)
\end{equation}    

A Walsh function $w_h$ can be implemented with a quantum operator $\hat{w}_h$, called \emph{Walsh operator}. On $n$ qubits, the Walsh operator $\hat{w}_h$ is defined as \cite{Welch2014}:
\begin{equation}\label{eq:walsh_op}
    \hat{w}_h = \hat{Z}^{h_{0}} \otimes \hat{Z}^{h_1} \otimes \cdots \otimes \hat{Z}^{h_{n-1}}
\end{equation}
where $\hat{Z}$ is the $Z$-Pauli operator and $h_j$ is the $j$-th least significant bit of $h$, with:
\begin{equation}
    \hat{Z}^{h_j} = \begin{cases}
        \hat{Z} & \text{ if } h_j = 1 \\
        \hat{I} & \text{ if } h_j = 0
    \end{cases}
\end{equation}
We adopt the convention that the most significant qubit is the leftmost one. 
Notice that operator in Equation \ref{eq:walsh_op} which is applied to the most significant qubit is elevated to the \emph{least significant bit} of $h$.
Of particular importance for our contribution is to highlight that the Walsh operator of order $h=0$ corresponds to the identity.

The Walsh operator in Equation \ref{eq:walsh_op} is equivalent to a $\hat{CNOT}$ staircase, constructed as follows. First, given $h$ we define $q_p$ as the most significant qubit on which $\hat{w}_h$ applies a $\hat{Z}$ gate. Then, we define operator $\hat{A}_{h}$ as:
\begin{equation}
    \hat{A}_{h} = \prod_{i=0}^{p-1}\pars{\hat{CNOT}^i_p}^{h_i}
\end{equation}
where $\hat{CNOT}^i_p$ is a unitary, defined on $n$ qubits, which applies a $\hat{CNOT}$ gate, controlled by $q_i$ with target $q_p$. Then, the Walsh operator $\hat{w}_h$ can be written as:
\begin{equation} \label{eq:new_walsh_op}
    \hat{w}_h = \hat{A}_{h}\pars{\bigotimes_{i=0}^{p-1}\hat{I} \otimes \hat{Z} \otimes \bigotimes_{i=p+1}^{n} \hat{I}}\hat{A}^{-1}_{h}
\end{equation}
\begin{figure}[ht]
    \centering
    \includegraphics[width=0.70\columnwidth]{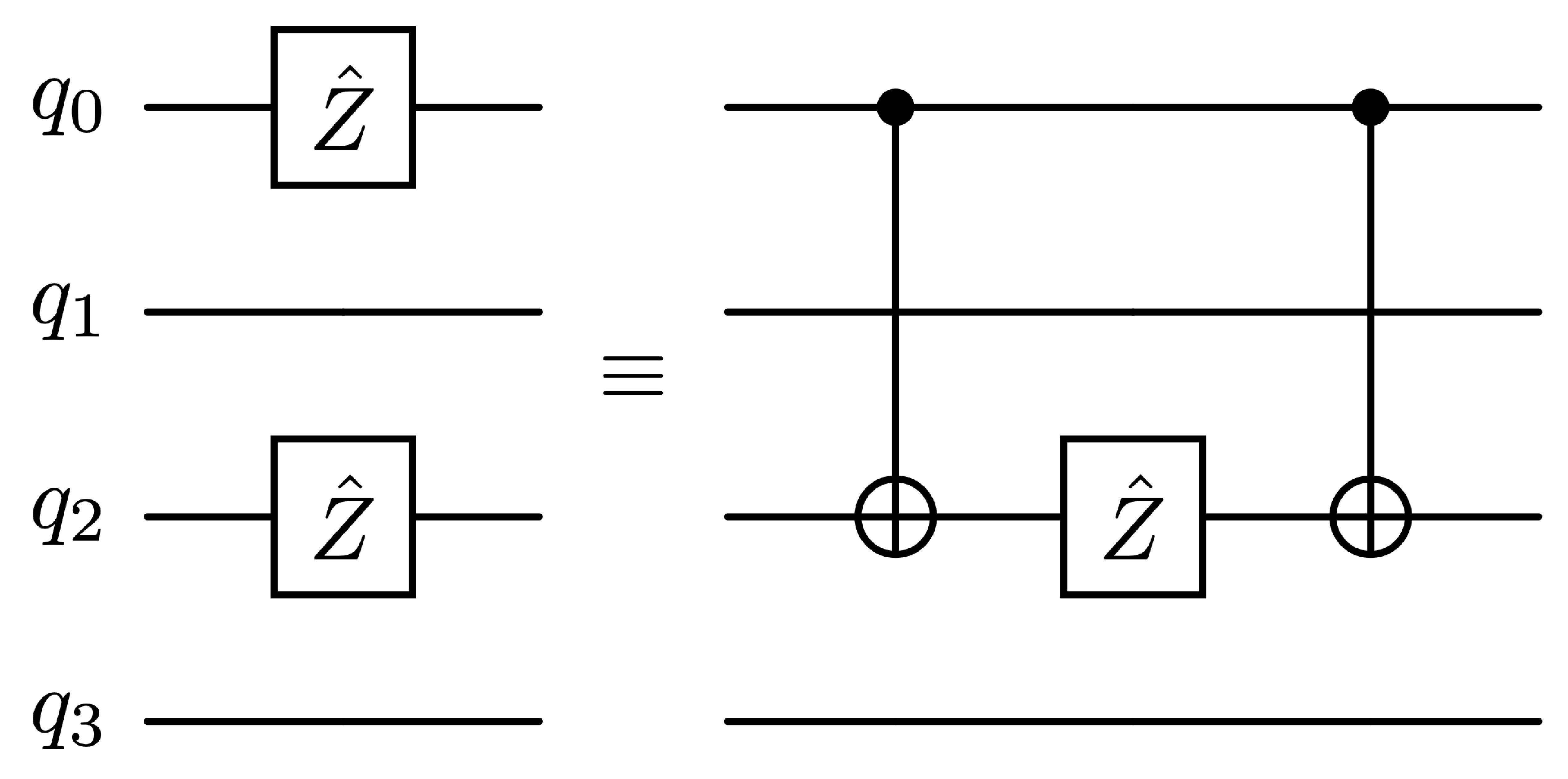}
    \caption{Equivalence between a layer of $\hat{Z}$ gates and the $\hat{CNOT}$ staircase for $h=10$, hence with $h_0=0, h_1=1, h_2=0, h_3=1$.}
    \label{fig:walsh_op_equivalences}
\end{figure}
Figure \ref{fig:walsh_op_equivalences} shows an example of $\hat{w}_h$.

Based on this it is straightforward to determine how to implement the operator $\hat{W}_h(a_h)$ as a quantum circuit for $h>0$, see Equation \ref{eq:W_h_circuit}. Notice that gate $\hat{Z}$ becomes unitary $e^{-ia_h\hat{Z}}$, corresponding to a $\hat{R}_z(2 a_h)$ gate. 

\begin{equation}
    \label{eq:W_h_circuit}
    \hat{W}_h(a_h) = \hat{A}_{h}\pars{\bigotimes_{i=0}^{p-1}\hat{I} \otimes \hat{R}_z(2a_h) \otimes \bigotimes_{i=p+1}^{n} \hat{I}}\hat{A}^{-1}_{h}
\end{equation}

At this point, however, we encounter a crucial issue. The $\hat{W}_0$ operator produces a global phase which is not possible to implement with a $\hat{R}_z(\theta)$ gate. This means that implementing $\hat{W}_0$ requires an ad-hoc strategy. 
\begin{align}
\hat{W}_0 &= e^{-i a_0 \hat{w}_0} 
= e^{-i a_0 \bigotimes^{n-1}_{j=0}\hat{I}_j}
= e^{-i a_0}
\label{eq:walsh_0}
\end{align}

If the Walsh operator is used in a non-controlled fashion, the missing term of order zero is not an issue because it would only produce a global phase that can be discarded. Hence there is no need to include it.

On the other hand, the WSL algorithm applies the operator in a controlled way, which means that the phase it generates is \emph{relative} on the ancilla qubit and therefore cannot be ignored. However, the description of the original WSL algorithm does not discuss this special case and does not describe how to include the term into the quantum circuit. In the following section we discuss how this issue can be addressed.

\begin{figure*}[hbt]
\centering
\begin{adjustbox}{width=0.85\textwidth}
\begin{quantikz}
    \alpha \quad \ket{0}  
        & \gate{\hat{H}} 
        & \ctrl{1} 
        & \ctrl{1} 
        & \ \ldots\ 
        & \ctrl{1}\slice{1} 
        & \gate{\underline{\hat{P}(-\epsilon_0 a_0)}}\slice{2} 
        & \gate{\hat{H}} 
        & \gate{\hat{P}\left( -\frac{\pi}{2} \right)} \slice{3} 
        & \meter{} \\
    q_0 \quad \ket{0} 
        & \gate{\hat{H}} 
        & \gate[3]{\hat{W}_1(\epsilon_0 a_1)} 
        & \gate[3]{\hat{W}_2(\epsilon_0 a_2)} 
        & \ \ldots\ & \gate[3]{\hat{W}_{M-1}(\epsilon_0 a_{M-1})} 
        & 
        & 
        & 
        &\\
    \vdots \\
    q_n \quad \ket{0} 
        & \gate{\hat{H}} 
        & 
        & 
        & \ \ldots\ 
        & 
        & 
        & 
        & 
        &
\end{quantikz}
\end{adjustbox}
\caption{Quantum circuit of the WSL algorithm. The least significant qubit is on the top of the circuit. $\hat{H}$ identifies an Hadamard gate, the operator $\hat{W}_h(\epsilon_0 a_h)$ identifies a term of the Walsh series of $f$, while $\hat{P}(\beta)$ identifies a phase gate that introduces a phase $\beta$. Note that gate $\hat{P}(-\epsilon_0 a_0)$ is underlined to highlight we added it as part of our contribution.}
\label{fig:wsl_circuit}
\end{figure*}
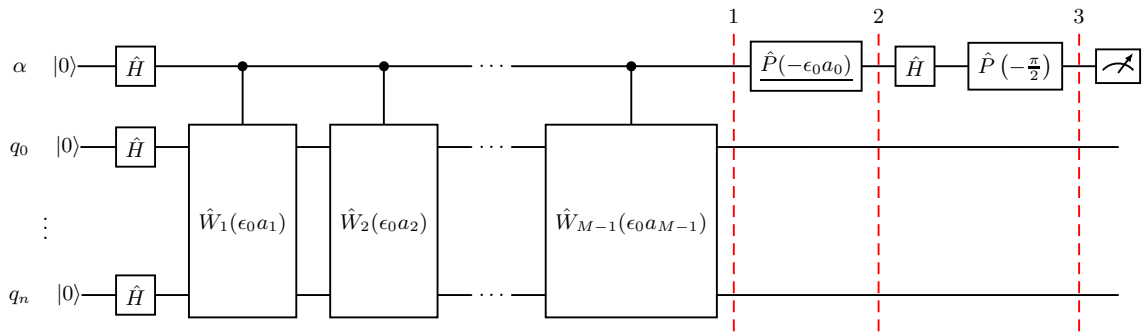

\subsection{Description of the Correct Walsh Series Loader}\label{subsec:imple_wsl}

The Walsh Series Loader performs state preparation by applying the Walsh operator $\hat{U}_{f}$ on a register of $n$ qubits, as a unitary controlled by an ancilla qubit $\alpha$ in superposition. The algorithm is a repeat-until-success method, where $\ket{f}$ has been prepared correctly only if the ancilla qubit is $\ket{1}$. The WSL algorithm introduces a coefficient $\epsilon_0 > 0$ multiplied to the Walsh transform $a_h$ to control the error on the prepared state as well as the probability of success. As $\epsilon_0 \to 0$ the fidelity of the prepared state increases. 
Given the discrete function $f$ defined over $N$ points, the state we aim to prepare is:
\begin{equation}
    \myket{f} = \frac{1}{\sqrt{\sum_{k=0}^{N-1}f(k)^2}}\sum_{k=0}^{N-1} f(k) \myket{k}
\end{equation}

The quantum circuit of the WSL algorithm is composed by a quantum register of $n$ qubits where we wish to prepare state $\ket{f}$ and one ancilla qubit. 
The first step is to compute the $N$ Walsh functions $w_h(k)$ and the Walsh transform $a_h$ of $f$. For a truncated series of $M<N$ terms, we can apply Equation \ref{eq:Walsh_transform}.

\begin{figure*}[hbt]
    \centering
    \includegraphics[width=0.75\textwidth]{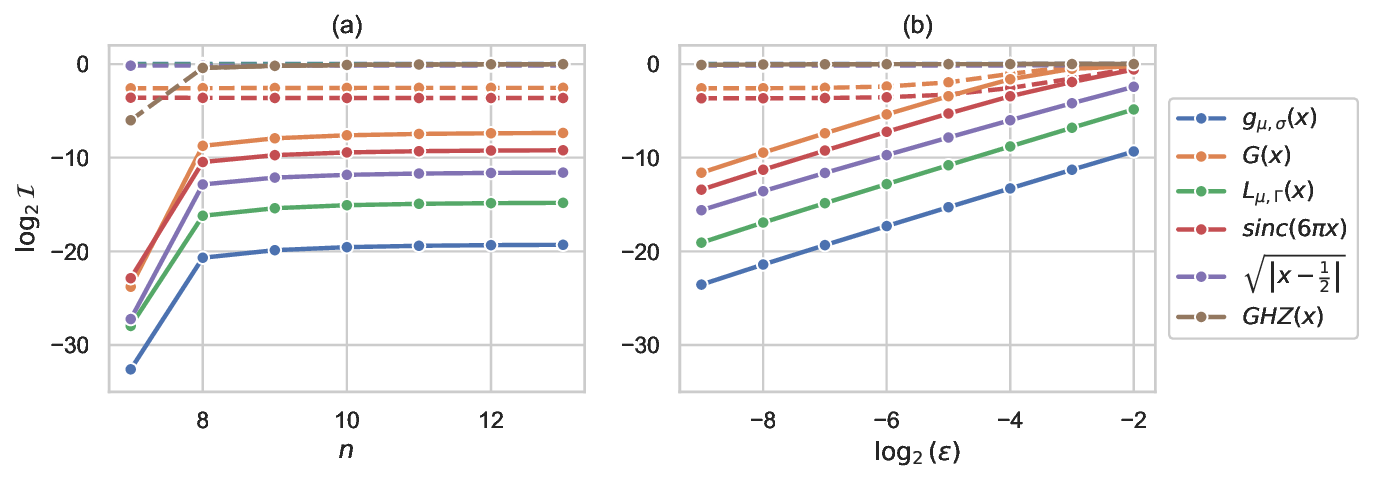}
    \caption{(a) Relationship between the number of qubits and the logarithm of the infidelity of the prepared state, lower infidelities are better. As in the original paper, we set $\epsilon_0 = \epsilon_1 = 2^{-7}$. Note that the infidelity for the $GHZ(x)$ on $7$ qubits is missing because it is $-\infty$.
    (b) Relationship between the logarithm of $\epsilon = \epsilon_0 = \epsilon_1$ and the logarithm of the infidelity of the prepared state, for a register of $12$ qubits.
    For both plots, dashed lines represent states prepared with the 
    incomplete WSL algorithm, while solid lines represent states prepared with the correct WSL described in this paper.}
    \label{fig:infid}
\end{figure*}

The second step is to create the quantum circuit shown in Figure \ref{fig:wsl_circuit}. 
Notice how every operator $\hat{W}_h(a_h \epsilon_0)$ is controlled by the ancilla qubit $\alpha$. 
At the step labeled with 1 in Figure \ref{fig:wsl_circuit}, the state is:
\begin{equation}
    \ket{\psi_1} = \frac{1}{\sqrt{2}}\left(\ket{s_0}\ket{0} + \ket{s_1}\ket{1}\right)
\end{equation}
where $\ket{s_0}$ is not relevant because the state is prepared correctly only of the ancilla is equal to $\ket{1}$, while $\ket{s_1}$ is:
\begin{equation}\label{eq:pre_phase_gate_state1}
    \ket{s_1} = \frac{1}{\sqrt{N}}\sum_{k=0}^{N-1}\exp\left(-i\epsilon_0
    \sum_{h=1}^{M-1} a_h w_h(k)
    \right)\ket{k}
\end{equation}
where the innermost summation corresponds to the Walsh terms of order $h\geq 1$. Notice that the Walsh term of order $h=0$ is missing because, as we discussed, it cannot be implemented in the same way as the other terms. Since the Walsh operators are applied as controlled unitaries by an ancilla, the global phase generated by $\hat{W}_0(a_0\epsilon_0)$ will end up as a relative phase on the ancilla and, as such, it cannot be ignored. 

In order to address this issue and include $\hat{W}_0(a_0\epsilon_0)$ we must add a phase $e^{-i a_0\epsilon_0}$ to all states of the quantum register, which corresponds to a relative phase applied to state $\ket{1}$ of the ancilla qubit. We achieve this by adding a phase gate $\hat{P}(-\epsilon_0 a_0)$ on the ancilla qubit immediately after the last controlled Walsh operator, see Figure \ref{fig:wsl_circuit}.

With our integration, this correct WSL circuit can work as intended. At the 
end of the circuit, for small enough $\epsilon_0$, state $\ket{1}$ of the ancilla qubit is entangled with a state of the quantum register asymptotic to $\ket{f}$. Since integrating this term only requires one gate, it has a negligible impact on the scaling of WSL.

\section{\label{sec:experiments}Experimental Analysis }
In this section we aim to show that if the Walsh term $\hat{W}_0(a_0\epsilon_0)$ is not correctly included in the implementation of WSL, the fidelity of the prepared state is very poor. This highlights the importance of this apparently small component.

\subsection{Experimental Setup}
In order to assess the fidelity of the states prepared with WSL we use the same functions $f$ as in the original paper. Notice that all functions are defined as continuous over the interval $[0,1)$, then are discretized on $N=2^n$ points, with $n=\{7,8,9,10,11,12,13\}$ qubits.

The functions are:
(i) Gaussian $
        g_{\mu, \sigma}(x) = \frac{1}{\sigma}\exp\left(-\frac{1}{2\sigma^2}(x-\mu)^2\right)
        $ with $\mu = 0.5$ and $\sigma=1$; 
(ii) bimodal Gaussian $
        G(x) = sg_{\mu_1, \sigma_1}(x) + (1-s)g_{\mu_2, \sigma_2}(x)
    $ with $\mu_1 = 0.25$, $\mu_2 = 0.75$, $\sigma_1 = 0.3$, $\sigma_2 = 0.04$ and $s = 0.1$;
(iii) Lorentzian: $
        L_{\mu, \Gamma}(x) = \frac{\Gamma}{\Gamma^2 + 4(x - \mu)^2}
    $ with $\Gamma = 1$ and $\mu = 0.5$.
(iv) cardinal sine:\footnote{Note that while the original paper states that the function has argument $6\pi x$, it reports a plot of a cardinal sine but with argument $19\pi x$.} $
        \text{sinc}(x) = \frac{sin(6\pi x)}{6 \pi x}
    $;
(v) a non-differentiable function $
        f(x)=\sqrt{\left| x - \frac{1}{2} \right|}
    $

To provide a more complete picture of WSL, we also include a Greenberger-Horne-Zeilinger state on $n$ qubits, which is maximally entangled and already discrete:
    \begin{equation*}
        GHZ(x) = \begin{cases}
            0 & \text{ if } x \in \left(0,\frac{N-1}{N}\right) \\
            \frac{1}{\sqrt{2}} & \text{ otherwise} 
        \end{cases}
    \end{equation*}

As in the original paper, we set $\epsilon_0 = \epsilon_1 = 2^{-7}$, which correspond to $M = 2^7$ Walsh terms.

The prepared state $\ket{\tilde{f}}$ is evaluated based on its \emph{infidelity} $\mathcal{I}$ with respect to the target state $\myket{f}$, defined as:
\begin{equation*}
    \mathcal{I}(\tilde{f}, f) = 1 - |\braket{\tilde{f}|f}|^2
\end{equation*}

\subsection{\label{subsec:results}Results}
In this section we compare the results of an 
incomplete WSL implementation that is missing term $\hat{W}_0(a_0\epsilon_0)$, with a correct one including it as described in this paper. As in the original paper, we report the logarithm of the infidelity of the prepared state which we show in relation to $n$ and to the value of $\epsilon=\epsilon_0=\epsilon_1$ (see Figure \ref{fig:infid}).

Plot (a) in Figure \ref{fig:infid} shows the relationship between the logarithm of the infidelity of the prepared state and the number of qubits. We notice that, given a state to prepare $\ket{f}$, the  
incomplete WSL algorithm exhibits an infidelity that is orders of magnitude worse compared to that obtained with the correct WSL algorithm we describe. Furthermore, the results obtained using the correct WSL implementation are comparable to those shown in the original article highlighting how the original results are reproducible only by addressing this special case.

Plot (b) shows the relationship between the infidelity of the prepared state and $\epsilon$, which is linear for the correct WSL algorithm. Again, the incomplete
WSL algorithm does not allow to reproduce the original results, but the correct version does. Indeed, the incomplete WSL algorithm also numerically violates the theoretical upper bound $\mathcal{I}(\tilde{f}, f)\leq \epsilon$ that is proven in the original paper, which are instead consistent with the correct WSL.
Lastly, Theorem 1 of the original WSL paper \cite{Zylberman2024} showed that truncated WSL may not prepare well highly entangled states. Indeed, operator $\hat{W}_h$ does not introduce entanglement. We can observe the effect empirically on the Greenberger-Horne-Zeilinger state which is prepared with reasonable fidelity only when $M = N$.

\section{Conclusions}
In this paper, we have discussed how the implementation of the algorithm described in the paper ``Efficient quantum state preparation with Walsh series" \cite{Zylberman2024} does not mention that the Walsh term operator $\hat{W}_0$ is as special case that cannot be implemented like the others. Without this term the results of the algorithm are very poor and is not possible to reproduce the results reported in the original paper.
We have shown how the missing term can be included by adding a phase gate on the ancilla qubit, and that this complete algorithm behaves as expected in preparing states with reasonable fidelity.
Finally, we show empirically the known theoretical limitations of WSL in preparing highly entangled states.

\section{Acknowledgment}
We acknowledge the financial support from ICSC - ``National Research Centre in High Performance Computing, Big Data and Quantum Computing'', funded by European Union - NextGenerationEU.

\clearpage
\appendix

\end{document}